\let\oldsqrt\sqrt
\def\sqrt{\mathpalette\DHLhksqrt}
\def\DHLhksqrt#1#2{%
\setbox0=\hbox{$#1\oldsqrt{#2\,}$}\dimen0=\ht0
\advance\dimen0-0.2\ht0
\setbox2=\hbox{\vrule height\ht0 depth -\dimen0}%
{\box0\lower0.4pt\box2}}

\documentclass[12pt,preprint]{aastex}      

\usepackage[latin1]{inputenc}
\usepackage{lscape}

\begin{document}

\title{Observing the onset of outflow collimation in a massive protostar}

\author{C. Carrasco-Gonz\'alez,\altaffilmark{1} J.M. Torrelles,\altaffilmark{2} J. Cant\'o,\altaffilmark{3} S. Curiel,\altaffilmark{3} G. Surcis,\altaffilmark{4}\\W.H.T. Vlemmings,\altaffilmark{5} H.J. van Langevelde,\altaffilmark{4,6} C. Goddi,\altaffilmark{4,7} G. Anglada,\altaffilmark{8}\\S.-W. Kim,\altaffilmark{9,10} J.-S. Kim,\altaffilmark{11} J.F. G\'omez\altaffilmark{8}}

\altaffiltext{1}{Centro de Radioastronom\'{\i}a y Astrof\'{\i}sica UNAM, Apartado Postal 3-72 (Xangari), 58089 Morelia, Michoac\'an, Mexico; 
$^2$Institut de Ci\`encies de l'Espai (CSIC) and Institut de Ci\`encies del 
Cosmos (UB)/Institut d'Estudis Espacials de Catalunya, Carrer Mart\'{\i} i Franqu\`es 1, 08028 Barcelona, Spain; 
$^3$Instituto de Astronom\'{\i}a (UNAM), Apartado 70-264, 04510 Mexico D. F., Mexico; 
$^4$Joint Institute for VLBI in Europe, Postbus 2, 79990 AA Dwingeloo, Netherlands; 
$^5$Chalmers University of Technology, Onsala Space Observatory, SE-439 92 Onsala, Sweden; 
$^6$Sterrewacht Leiden, Leiden University, Postbus 9513, 2300 RA Leiden, Netherlands; 
$^7$Department of Astrophysics, Institute for Mathematics, Astrophysics and Particle Physics, Radboud University, Post Office Box 9010, 6500 GL Nijmegen, Netherlands; 
$^8$Instituto de Astrof\'{\i}sica de Andaluc\'{\i}a (CSIC), Apartado 3004, 18080 Granada, Spain;
$^9$Korea Astronomy and Space Science Institute, 776 Daedeokdaero, Yuseong, Daejeon 305-348, Republic of Korea; 
$^{10}$Korea University of Science and Technology, 217 Gajeong-ro Yuseong-gu, Daejeon 305-350, Republic of Korea; 
$^{11}$National Astronomical Observatory of Japan, 2-21-1 Osawa, Mitaka, Tokyo 181-8588, Japan}

{\fontfamily{qhv}\selectfont
The current paradigm of star formation through accretion disks, and magnetohydrodynamically driven gas ejections, predicts the development of collimated outflows, rather than expansion without any preferential direction. We present radio continuum observations of the massive protostar W75N(B)-VLA 2, showing that it is a thermal, collimated ionized wind and that it has evolved in 18 years from a compact source into an elongated one. This is consistent with the evolution of the associated expanding water-vapor maser shell, which changed from a nearly circular morphology, tracing an almost isotropic outflow, to an elliptical one outlining collimated motions. We model this behavior in terms of an episodic, short-lived, originally isotropic, ionized wind whose morphology evolves as it moves within a toroidal density stratification.}

\vspace{1cm}

Water-vapor masers at 22 GHz are commonly found in star-forming regions, arising in the shocked regions created by powerful outflows from protostars in their earliest phases of evolution (\emph{1}). Observations of these masers with very long baseline interferometry (VLBI) indicate that at the early life of massive stars there may exist episodic, short-lived  (tens of years) events associated with very poorly collimated outflows (\emph{2,3,4,5}). These results are surprising because, according to the core-accretion model for the formation of massive stars  ($\gtrsim$8~M$_{\odot}$), which is a scaled-up version of low-mass star formation, collimated outflows are already expected at their very early phases (\emph{6,7,8}). 

A unique case of a short-lived, poorly collimated outflow is the one found in the high-mass star-forming region W75N(B). This region contains two massive protostars, VLA~1 and VLA~2,  separated by $\lesssim$0.7 arc sec [projected separation of $\lesssim$910~astronomical units (AU) at the source distance of 1.3 kpc; (\emph{9})], both associated with strong water-vapor maser emission at 22 GHz (\emph{10,11}), and with a markedly different outflow geometry. At epoch 1996, VLA~1 shows an elongated radio continuum emission consistent with a thermal radio jet, as well as water maser emission tracing a collimated outflow of $\sim$1300~AU along its major axis. In contrast, in VLA~2 the water masers traced a shock-excited shell of $\sim$185~AU diameter radially expanding with respect to a central, compact radio continuum source ($\lesssim$0.12~arc sec, $\lesssim$160 AU) of unknown nature (\emph{12}). A monitoring of the water masers toward these two objects from 1996 to 2012 shows that the masers in VLA~1 display a persistent linear distribution along the major axis of the radio jet.  In the case of VLA~2 we observe that the water maser shell continues its expansion at $\sim$30~km~s$^{-1}$ 16 years after its first detection.  More important, the shell has evolved from an almost circular structure ($\sim$185~AU to an elliptical one ($\sim$354$\times$190~AU) oriented northeast-southwest, along a direction similar to that of the nearby VLA~1 radio jet (\emph{13,14})  (fig.~S1) and of the ordered large-scale (2000 AU) magnetic field observed in the region (\emph{15}). The estimated kinematic age for the expanding shell is $\sim$25~years, indicating that it is driven by a short-lived, episodic outflow event. Moreover, our polarization measurements of the water maser emission show that whereas the magnetic field around VLA 1 has not changed over time, the magnetic field around VLA~2 changed its orientation following the direction of the major axis of the water maser elliptical structure. That is, it now shares a similar northeast-southwest orientation with those of both the magnetic field around VLA~1 and the large-scale magnetic field in the region (\emph{14,15})(fig.~S1). 

All these observations suggest that we are observing in ``real time'' the transition from an uncollimated outflow to a collimated outflow during the early life of a massive star. This scenario predicts that, within the same time span of the evolution of the shell, VLA~2 must have also evolved from a compact radio continuum source to an extended elongated source along the major axis of the water maser shell. Furthermore, the radio continuum emission of VLA~2 should have physical properties (e.g., spectral energy distribution, size of the source as a function of frequency) characteristic of free-free emission from a thermal, collimated ionized wind (\emph{16,17}). This can be tested through continuum observations at centimeter wavelengths that usually traces the emission from collimated, ionized winds (\emph{17,18} ).

Taking advantage of the high sensitivity and high angular resolution of the Jansky Very Large Array (VLA) at centimeter wavelengths, we obtained new observations in 2014 at several bands in the frequency range from 4 to 48 GHz (\emph{19}). These highly sensitive observations confirmed the expected scenario proposed above. The source VLA 2 is detected at all bands. In the images of higher frequency bands [U ($\sim$15~GHz), K ($\sim$23~GHz), and Q ($\sim$44~GHz)], which have higher angular resolutions ($\sim$0.1 to 0.2~arc sec), the source VLA~2 appears clearly elongated in the northeast-southwest direction (Fig.~1). Water maser emission was also observed simultaneously with the K band continuum emission (\emph{19}), allowing a very accurate alignment (better than $\sim$1~milli-arc sec) between the masers and the continuum emission. We find that the elongation of the continuum emission is in good agreement with that of the water maser distribution (Fig.~1).

Comparison of the radio continuum emission of VLA~2 at K band between epochs 1996 ({\emph{10}) and 2014 is shown in Fig.~2. Whereas in 1996 the emission was compact, in 2014 we observed extended emission in the northeast-southwest direction. In particular, the core of the radio continuum emission of VLA~2 has evolved from a compact source in 1996  ($\lesssim$160~AU), to an elongated core with a full width at half maximum (FWHM) of $\sim$220~AU$\times$$\lesssim$160~AU (position angle $\simeq$65$^{\circ}$) in 2014.  This evolution is consistent with those of the expanding water maser shell and of the magnetic field, over the same time span (fig.~S1) (\emph{14}). This core elongation is clearly seen in all the images that we have made by weighting the interferometer uv data in different ways, from natural to uniform (\emph{19}) (fig.~S2). In addition, we also found that the VLA~2 source exhibits weak extended emission at distances of $\sim$390~AU southwest from the peak position, which was not detected in 1996. Because the sensitivity attained in our 2014 observations ($\sim$10~$\mu$Jy~beam$^{-1}$) is a factor of $\sim$15 better than in 1996 ($\sim$150~$\mu$Jy~beam$^{-1}$), it is plausible that the weak extended emission was already present in 1996, but not detected because of the lower sensitivity in those observations. That the peak intensity of VLA~2 has significantly decreased from 1996 to 2014, while total flux densities in both epochs have remained similar (table S3), rather suggests that the radio continuum emission from VLA~2 arises from a larger, more elongated structure in the 2014 epoch (see model below). In addition, the distribution of the water masers follows the observed morphology of the continuum emission at both epochs (Fig.~2), indicating that they trace a molecular shell surrounding the radio continuum source. We also considered the possibility that the extended radio continuum emission in 2014 is actually the signature of a new variable protostar. Nevertheless, this hypothesis is rendered unlikely by the spatial coincidence of this new radio continuum emission with VLA 2.

As a result of  the wide bandwidths provided by the VLA, we performed an analysis of the behavior of the flux density and of the size of the major axis of VLA~2 with frequency (\emph{19}) (table S3 and fig.~S4). This analysis was carried out only at U, K, and Q bands, whose high angular resolution allows us to completely isolate the emission of VLA~2 from the other sources in the field. We find that the flux density ($S_\nu$) at the lowest observed frequencies (U and K bands) is dominated by free-free emission with a spectral index $\alpha_{ff}\simeq$0.61 ($S_\nu\propto\nu^{\alpha_{ff}}$). Moreover, we also find that the size of the major axis of VLA~2 ($\theta$) decreases with frequency as a power law ($\theta \propto \nu^\beta$), with an exponent $\beta\simeq-$0.7. The behavior of both the flux density and the size of the major axis of VLA~2 is in good agreement with what is expected from a thermal, collimated and ionized wind (\emph{16,17}). The presence of water maser emission and the detection of shock-excited SiO emission highly concentrated towards this source (\emph{20}) further supports a very fast wind interaction region rather than an HII region. The VLA~2 wind could contribute, together with the winds from the other young stellar objects (YSOs) in the region, to the large-scale ($\sim$2~pc) molecular outflow observed in the W75N(B) star-forming region (\emph{21}).

 At the highest observed frequencies (Q band), the emission deviates by $\sim$30$\%$ from the expected flux densities of the ionized wind at these frequencies, showing in addition a steeper spectral index  that can be easily explained by the presence of thermal dust emission. The presence of thermal dust emission at frequencies higher than 40~GHz is commonly found in protostars and is related to the presence of a dusty circumstellar disk around the protostar and/or to a dusty envelope.  Our data suggest that the spectral index of the dust component in the observed wavelength range is $\alpha_{dust}$ $\simeq$ 3 to 4 (fig.~S4), a value similar to what is commonly found in massive protostars (\emph{22,23}). 
 
In summary, our observations reveal that the radio continuum source VLA~2 has evolved in only 18 years from being a compact source, to become an elongated source in the same direction as that of the water maser shell and the magnetic field. Moreover, the characteristics of the continuum emission and the distribution of the water masers in VLA~2 are now consistent with a thermal, collimated ionized wind, surrounded by a dusty disk or envelope. We interpret these results in terms of the evolution of an episodic, short lived, originally isotropic ionized wind interacting with a toroidal environmental density stratification.

The interaction of a wind with the surrounding environment produces a two-shock structure that travels away from the star. The outer shock accelerates the environment, whereas the inner shock decelerates the wind. For young stars with moderate wind velocities (few hundred km~s$^{-1}$), both the inner and outer shocks are, in general, radiative and result in the formation of a thin expanding shell bounded by the two shocks. In (\emph{24}) we model this scenario for the case of a steady isotropic wind with mass loss rate ${\dot M_w}$ and terminal velocity $V_w$, surrounded by a torus of dense material with a density distribution of the form $\rho(R,\theta) = \rho_0/(1+[R/R_c (\theta)]^2)$. $R$ is the distance from the star; $R_c(\theta) = R_0(1 + p \sin^2\theta)$ is the latitude-dependent radial size of the core, with $\theta$ the polar angle measured from the symmetry axis of the torus; $\rho_0$ is the density of the molecular core; and $p$ is related to the density contrast between the equator ($\theta = \pi/2$) and the pole ($\theta = 0$) of the toroidal environment. This density distribution implies that for  $R \ll R_c$ the density is nearly uniform, that is, independent of direction and radial size, whereas for $R \gg R_c$ the density decreases as 1/$R^2$, which is the behavior of the density of self-gravitating isothermal spheres at large distances. 

Our model (\emph{24}) shows that at small radii ($R \ll R_c$ ) the shell grows isotropically (with no angular dependence), the initial velocity of the expanding shell is the velocity of the wind, and the shell decelerates owing to the incorporation of environmental material with zero velocity. The expansion at large radii ($R \gg R_c$) has two important physical characteristics. First, there is a strong angular dependence of the expansion velocity of the shell, being much larger along the symmetry axis of the torus than along its equator; and second, this expansion velocity 
asymptotically tends to an angular-dependent constant value. Thus, this isotropic wind + torus model can explain the distinctive characteristics of our radio continuum and water maser observations (the transition from a nearly spherical outflow to a collimated, elongated one over a short period of time, and the expansion of the masers at constant velocity), because the shell tends to expand faster along the symmetry axis of the surrounding torus (Fig.~3).

In our model (\emph{24}; Fig.~3), we identify the semi-major axis of the water maser shell reported at different epochs by (\emph{14}) with the evolution of the shell radius along the symmetry axis of the surrounding torus ($\theta = 0$), and the semi-minor axis of the water maser shell with the shell size along the equator ($\theta = \pi/2$). Taking this into account, and assuming that the radio continuum source VLA 2 traces the emission of the ionized wind that drives the shell, we can find a range of possible physical values to fit the observations. In particular, we obtain values for the mass loss rate and terminal wind velocity of $\dot M_w \simeq$ (4-11)$\times10^{-7}$~M$_\odot$ yr$^{-1}$ and $V_w \simeq$ 110 to 350~km~s$^{-1}$, respectively. In addition, for the initial epoch of ejection of the wind we found $t_i \simeq$ 1984 to 1985, with uniform density core radial size $R_0 \simeq$ 26 to 29~AU and particle density $n_0 \simeq$ (4 to 6)$\times10^7$~cm$^{-3}$. 

All the parameters estimated by our model match those expected for a massive protostar embedded in a high-density environment. Furthermore, from the thermal dust emission at 7~mm (after removal of the free-free contamination) we roughly estimate a total (dust+gas) mass of $\gtrsim$ 0.001~M$_{\odot}$ (\emph{19}). This mass corresponds to an average particle density of $\gtrsim$2$\times 10^7$~cm$^{-3}$ in a region of $\sim$0.2~arc sec (the maximum size of the region where we detect 7~mm continuum emission; Fig.~1). Within the uncertainties of the estimates (\emph{19}), we find that these values are consistent with the parameters of the toroidal gas structure.

Our data do not allow us to determine what the morphology of the radio continuum emission was like at scales $\lesssim$0.1 arc sec in 1996. In particular, we cannot rule out that the source was already slightly elongated with a size smaller than the beam in 1996. However, 1996 corresponds to an epoch $\sim$10 years after the estimated launch time (1983 to 1984). According to our model, with that time interval, some deviation from an originally spherical wind should already be expected (Fig.~3). 
The combination of all our radio continuum results, together with other important available data, can provide a coherent scenario for this source. Our scenario proposes a change in the collimation of the wind, from a poorly collimated outflow (maybe initially spherical) to a highly collimated wind, and this is consistent with all the available data: a compact continuum source and water masers poorly collimated in 1996, an elongated radio continuum source and collimated water masers in 2014. We can then affirm that a change in the collimation of the wind has taken place in this time span. We also find that the ionized wind, the water maser shell, and the magnetic field in VLA~2 (at scales of few hundred AU) have all evolved spatially in the same sense, sharing now a similar northeast-southwest orientation. This is similar to the orientation of the magnetic field in the neighboring radio jet VLA~1 and of the ``large-scale'' ($\sim$2000~AU) magnetic field in the region (\emph{14,15}). Therefore, the magnetic field at these ``large scales'' might control the formation of both objects (VLA~1 and VLA~2). In particular,  we think that the magnetic field could have favored the formation of the toroidal gas structure around VLA~2, with its symmetry axis also oriented northeast-southwest, as in our model.

The proposed scenario can thus explain the singular evolution of an episodic, short-lived ($\sim$30~years), originally non-collimated outflow, traced by the water masers, into a collimated outflow by its interaction with the ambient medium. Our observations and modeling reveal that the collimation of these short-lived outflow events from massive protostars occurs at relatively large distances from the central star, at radii between $\sim$30~AU (the radial size obtained from our model for the initial expansion of the shell) and $\sim$100~AU (the radius where the expanding shell turned out from a round to an elliptical shape; fig.~S1) [see also (\emph{25,26})]. These scales are on the order of those predicted for poorly collimated outflows in magnetized, massive collapsing cores during the very early stages ($\sim$10$^3$ to 10$^4$~years), according to the magnetohydrodynamic simulations developed by (\emph{27}). However, these simulations can only produce relatively low-velocity ($\sim$10~km~s$^{-1}$) poorly collimated outflows, which seem still insufficient to explain the larger wind velocities ($\gtrsim$100~km~s$^{-1}$) needed to drive the expanding motions of the shock-excited water maser shell in VLA~2. We also considered the scenario in which a very young spherical compact HII region embedded in an accretion disk begins to expand, following the model in (\emph{28}).  In this model, the HII region evolves to a bipolar morphology along the symmetry axis of the accretion disk, producing a wind driven by the thermal pressure of the ionized gas. However, the relatively low velocity of that wind ($\lesssim$30~km~s$^{-1}$) (\emph{26}) seems also insufficient to shock excite and drive the expanding water maser shell.

In summary, although episodic, short-lived outflows in massive protostars are probably related to episodic increases in the accretion rates, as observed in low-mass star formation (\emph{29}), the origin  of poorly collimated outflows with relatively high velocities, as observed in VLA~2, deserves further theoretical research. Our observations show the ``real time'' evolution of such an originally poorly collimated outflow into a collimated one. This opens a new exciting window of opportunity to study how the basic ingredients of star formation (e.g., molecular outflow, ionized wind, magnetic field) evolve over the next few years, providing insights that may have important implications for our knowledge of the early stages of high-mass star formation. We may be on the brink of describing and modeling in ``real time'' all of these rapid changes.

\begin{figure}
\begin{center}
\includegraphics[width=\textwidth]{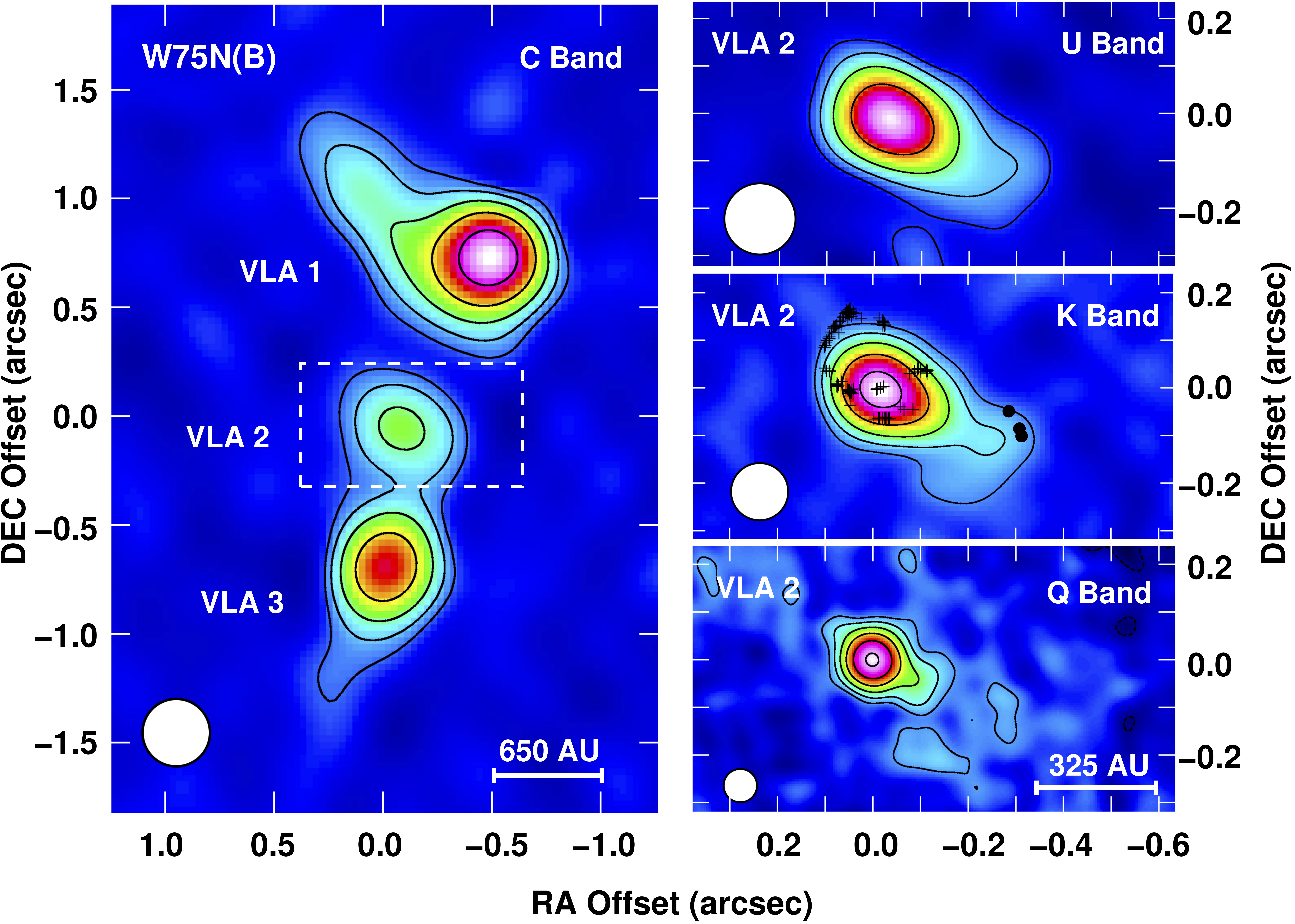}
\caption{\footnotesize{{\bf Radio continuum maps of the core of W75N(B) star-forming region obtained with the VLA in 2014 at several bands (C, U, K and Q)}. Left panel shows the C band (5~GHz) image of the region enclosing the sources VLA~1, 2, and 3. Right panels show close-ups to the VLA 2 source at U (15~GHz), K (23~GHz) and Q (43~GHz) bands. In the K band panel, we also show the positions of the water masers (plus symbols) observed simultaneously to the K band continuum, as well as the methanol masers (black dots). As seen in the right panels, VLA~2 appears elongated at all frequencies. This elongation is similar to that of the maser distribution. Contours are 4, 8, 16, 32, 64, and 128 times the root mean square (RMS) of each map: 30~$\mu$Jy~beam$^{-1}$ (C band), 11~$\mu$Jy~beam$^{-1}$ (U band), 10~$\mu$Jy~beam$^{-1}$ (K band), and 12~$\mu$Jy~beam$^{-1}$ (Q band). The restoring circular beam of each map (shown in the bottom left corner) is 0.31$''$, 0.15$''$, 0.12$''$, and 0.07$''$ for C, U, K, and Q band, respectively (0.1$''\simeq$ 130~AU at the source distance of 1.3~kpc). In all panels, the absolute position of the (0,0) is at right ascension RA(J2000) = 20$^h$38$^m$36.486$^s$ and declination DEC(J2000) = $+$42$^{\circ}$37$'$34.09$''$ ($\pm$ 0.03$''$), the peak position of VLA~2 at Q band, where the massive protostar is expected to be located. The accuracy in the relative positions of the water and methanol masers with respect to the K band continuum is better than $\sim$1 and $\sim$30~milli-arc sec, respectively (\emph{19}).}}
\end{center}
\end{figure}

\begin{figure}
\begin{center}
\includegraphics[width=0.7\textwidth]{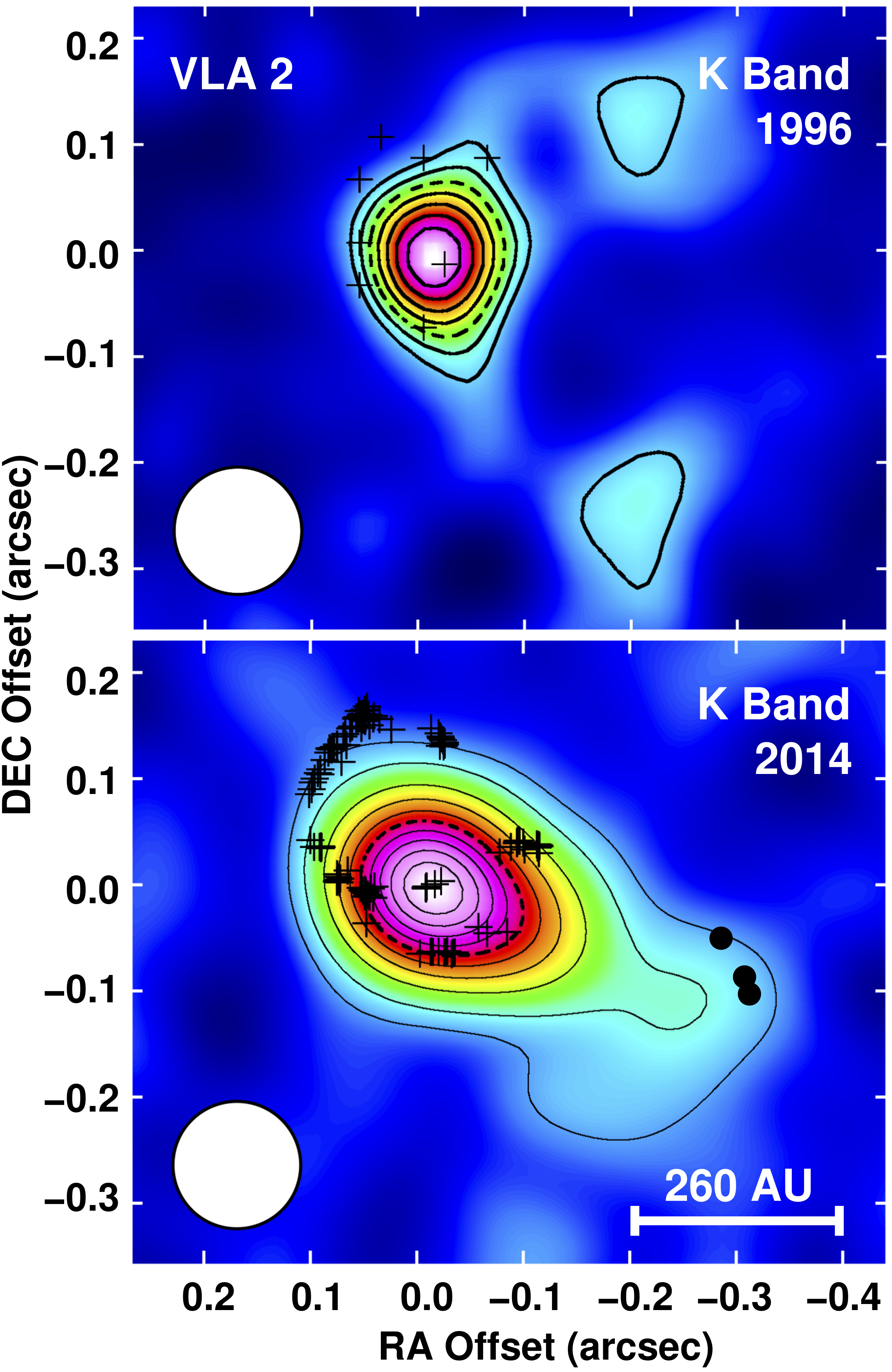}
\caption{\footnotesize{{\bf Comparison of the K band continuum emission of VLA~2 in epochs 1996 (top) and 2014 (bottom)}. The (0,0) position is the same as in Fig.~1. Contours are 30, 40, 50, 60, 70, 80, and 90\% of the peak intensity in 1996 (1.42~mJy~beam$^{-1}$) (RMS = 150~$\mu$Jy~beam$^{-1}$), and 5, 10, 20, 30, 40, 50, 60, 70, 80, and 90\% of the peak intensity in 2014 (0.82~mJy~beam$^{-1}$) (RMS = 10~$\mu$Jy~beam$^{-1}$). Both maps were obtained with the same restoring circular beam of 0.12$''$ (shown in the bottom left corner of each panel). In both panels, the half-power level is shown as a dashed line. We also show the water maser positions (plus symbols) for epochs 1996 and 2014 as observed with the VLA by (\emph{10}) and this work, respectively. The position of the methanol masers for epoch 2014 (black dots; this work) are also indicated. The FWHM size of the radio continuum emission has evolved from a compact source into an elongated source in the northeast-southwest direction, in a direction similar to that of the observed evolution of the water maser shell and magnetic field in VLA~2 (\emph{14}) (fig.~S1).}}
\end{center}
\end{figure}

\begin{figure}
\begin{center}
\includegraphics[width=1.0\textwidth]{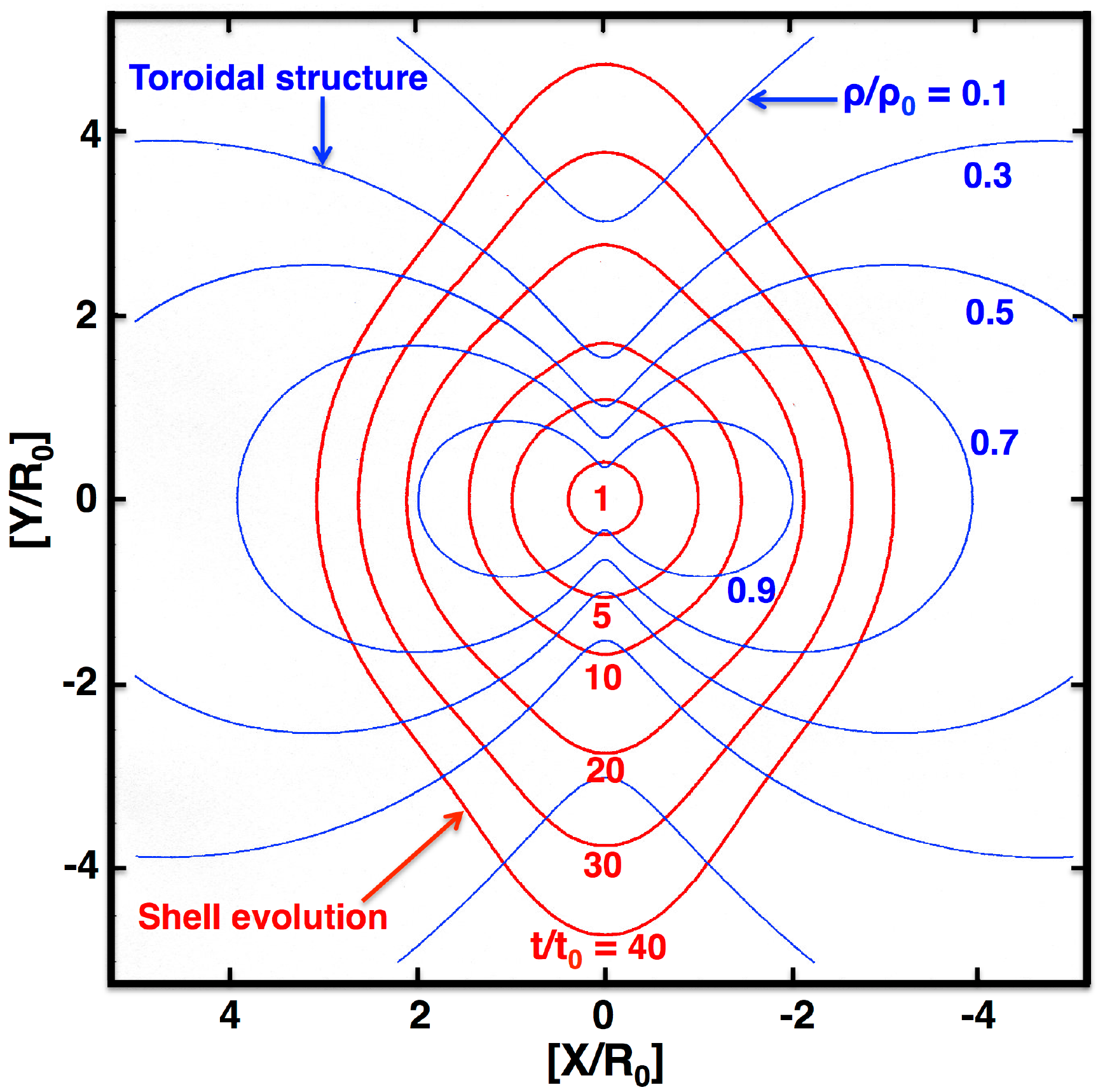}
\caption{\footnotesize{{\bf Evolution of an isotropic ionized wind driving the shock-excited water maser shell and that interacts with a toroidal environmental density stratification (\emph{24})}. The model predicts an initial round  shell, evolving into an elongated shell that expands faster along the symmetry axis of the  torus. Here we represent density isocontours of the toroidal environment (blue contours) for values $\rho/\rho_0$ = 0.1, 0.3, 0.5, 0.7, and 0.9 [eqs. E1 and E2 with p = 5; (\emph{24})]. The spatial evolution of the shell as a function of time $t$ after the initial ejection of the wind (red contours) is shown for values $t/t_0$ =  1, 5, 10, 20, 30, and 40 with $t_0$ = 0.5 years [adopting $V_w$ = 250~km~s$^{-1}$ and $R_0$ = 28~AU from (\emph{24})].}}
\end{center}
\end{figure}

\newpage

\begin{center}
\textbf{REFERENCES AND NOTES}
\end{center}

\begin{itemize}

\item{} 1. D. Hollenbach, M. Elitzur, C. F. McKee, Interstellar H$_2$O masers from J shocks. \emph{Astrophys. J.} \textbf{773}, 70 (2013)

\item{} 2. J. M. Torrelles et al., Spherical episodic ejection of material from a young star.  \emph{Nature} \textbf{411}, 277 (2001)

\item{} 3. G. Surcis et al., The structure of the magnetic field in the massive 
star-forming region W75N. \emph{Astron. Astrophys.} \textbf{527}, A48 (2011)

\item{} 4. J. M. Torrelles et al.,  A wide-angle outflow with the simultaneous presence of a high-velocity jet in the high-mass Cepheus A HW2 system.  
\emph{Mon. Not. R. Astron. Soc.} \textbf{410}, 627 (2011)

\item{} 5. L. Moscadelli et al., Massive star-formation in G24.78+0.08 explored through VLBI maser observations. 
\emph{Astron. Astrophys.} \textbf{472}, 867 (2007)

\item{} 6. G. Garay, S. Lizano,   Massive stars: their environment and formation.  \emph{Publications of the Astronomical Society of the Pacific} \textbf{111}, 1049 (1999)

\item{} 7. C. F. McKee, J. C. Tan,   The formation of massive stars from turbulent cores. \emph{Astrophys. J.}  \textbf{585}, 850 (2003)

\item{} 8. J. M. Girart, M. T. Beltr\'an, Q. Zhang, R. Rao, R. Estalella, Magnetic fields in the formation of massive stars. \emph{Science} \textbf{324}, 1408 (2009)

\item{} 9. K. L. J. Rygl et al.,   Parallaxes and proper motions of interstellar masers toward the Cygnus X star-forming complex. I. Membership of the Cygnus X region. \emph{Astron. Astrophys.}  \textbf{539}, A79 (2012)

\item{} 10. J. M. Torrelles et al., A radio jet-H$_2$O maser system in W75N(B) at a 200 au scale: exploring the evolutionary stages of young stellar objects. \emph{Astrophys. J.} \textbf{489}, 744 (1997)

\item{} 11. D. S. Shepherd, S. E. Kurtz, L. Testi,  The nature of the massive young stars in W75N.  \emph{Astrophys. J.} \textbf{601}, 952 (2004)

\item{} 12. J. M. Torrelles et al.,  Evidence for evolution of the outflow collimation in very young stellar objects.  \emph{Astrophys. J.} \textbf{598}, L115 (2003)

\item{} 13. J-S. Kim et al.,  Evolution of the water maser expanding shell in W75N VLA 2. \emph{Astrophys. J.} \textbf{767}, 86 (2013)

\item{} 14. G. Surcis et al.,   Rapidly increasing collimation and magnetic field changes of a protostellar H$_2$O maser outflow. \emph{Astron. Astrophys.} \textbf{565}, L8 (2014)

\item{} 15. G. Surcis, W. H. T. Vlemmings, R. Dodson, H. J. van Langevelde, Methanol masers probing the ordered magnetic field of W75N. \emph{A\&A} \textbf{506}, 757 (2009)

\item{} 16. S. P. Reynolds, Continuum spectra of collimated, ionized stellar winds. \emph{Astrophys. J.} \textbf{304}, 713 (1986)

\item{} 17. L. F. Rodr\'{\i}guez et al.,    Cepheus A HW2: A powerful thermal radio jet.  \emph{Astrophys. J.} \textbf{430}, L65 (1994)

\item{} 18. G. Anglada et al., Spectral indices of centimeter continuum sources in star-forming regions: implications on the nature of the outflow exciting sources. \emph{Astron. J.} \textbf{116}, 2953 (1998)

\item{} 19. See supplementary materials on Science Online

\item{} 20. Y. C. Minh, Y.-N. Su, H.-R. Chen, S.-Y. Liu, C.-H. Yan, S.-J. Kim, Submillimeter array observations toward the massive star-forming core MM1 of W75N. \emph{Astrophys. J.} \textbf{723}, 1231 (2010)

\item{} 21. D. S. Shepherd, L. Testi, L., D. P. Stark, Clustered star formation in W75~N. \emph{Astrophys. J.} \textbf{584}, 882 (2003)

\item{} 22. M. Osorio, S. Lizano, P. D'Alessio, Hot molecular cores and the formation of massive stars. \emph{Astrophys. J.} \textbf{525}, 808 (1999)

\item{} 23. L. T. Maud, M. G. Hoare, A. G. Gibb,  D. Shepherd, R. Indebetouw, High angular resolution millimetre continuum observations and modelling of S140-IRS1.
\emph{Mon. Not. R. Astron. Soc.} \textbf{428}, 609 (2013)

\item{} 24. See section E in supplementary materials on Science Online

\item{} 25. L. D. Matthews et al., A feature movie of SiO emission 20-100 AU from the massive young stellar object Orion source I. \emph{Astrophys. J.} \textbf{708}, 80 (2010)

\item{} 26.   L. J. Greenhill, C. Goddi,  C. J. Chandler,  L. D. Matthews, E. M. L. Humphreys,  Dynamical evidence for a magnetocentrifugal wind from a 20 M$_{\odot}$ binary young stellar object. \emph{Astrophys. J.} \textbf{770}, L32 (2013)

\item{} 27. D. Seifried, R. E. Pudritz, R. Banerjee,  D. Duffin, R. S. Klessen,  Magnetic fields during the early stages of massive star formation - II. A generalized outflow criterion. \emph{Mon. Not. R. Astron. Soc.} \textbf{422}, 347 (2012)

\item{} 28. E. Keto, The formation of massive stars: accretion, disks, and the development of hyper compact HII regions. \emph{Astrophys. J.} \textbf{666}, 976 (2007)

\item{} 29. J. M. Torrelles et al., A very young, compact bipolar H$_2$O maser outflow in the intermediate-mass 
star-forming LkH$\alpha$~234 region. \emph{Mon. Not. R. Astron. Soc.} \textbf{442}, 148 (2014)

\item{} 30. https://science.nrao.edu/facilities/vla/data-processing

\item{} 31. https://science.nrao.edu/facilities/vla/data-processing/pipeline

\item{} 32. A. Rau, T. J. Cornwell,   A multi-scale multi-frequency deconvolution algorithm for synthesis imaging in radio interferometry. \emph{Astron. Astrophys.} \textbf{532}, A71 (2011)

\item{} 33. L. S. G. Meehan, B. A. Wilking, M. J. Claussen, L. G. Mundy, A. Wootten,  Water masers in the circumstellar environments of young stellar objects. 
\emph{Astron. J.} \textbf{115}, 1599 (1998)

\item{} 34. P. D'Alessio, N. Calvet, L. Hartmann, Accretion disks around young objects. III. Grain growth. \emph{Astrophys. J.} \textbf{553}, 321 (2001)

\item{} 35. P. D'Alessio, N. Calvet, L. Hartmann, R. Franco-Hern\'andez, H. Serv\'{\i}n,
Effects of dust growth and settling in T Tauri disks. \emph{Astrophys. J.} \textbf{638}, 314 (2006)

\item{} 36. J. Cant\'o, A. C. Raga, L. Adame, The motion of wind-driven shells. \emph{Mon. Not. R. Astron. Soc.} \textbf{369}, 860 (2006)

\item{} 37. N. Panagia, M. Felli, The spectrum of the free-free radiation from extended envelopes. \emph{Astron. Astrophys.} \textbf{39}, 1 (1975)

\item{} 38. A. E. Wright, M. J. Barlow, The radio and infrared spectrum of early-type stars undergoing mass loss. \emph{Mon. Not. R. Astron. Soc.} \textbf{170}, 41 (1975)

\item{} 39. M. A. Trinidad et al., Formation and evolution of the water maser outflow event in AFGL 2591 VLA 3-N. 
\emph{Mon. Not. R. Astron. Soc.} \textbf{430}, 1309 (2013)

\item{} 40. G. Anglada, L. F. Rodr\'{\i}guez, C. Carrasco-Gonz\'alez, Radio jets in young stellar objects with the SKA.
In \emph{Proceedings of Advancing astrophysics with the Square Kilometre Array, PoS(AASKA14) 121}, 
arXiv:1412.640 (2014)

\item{} 41. I. Jim\'enez-Serra, J. Mart\'{\i}n-Pintado, A. B\'aez-Rubio, N. Patel, C. Thum, Extremely broad radio recombination maser lines toward the high-velocity ionized jet in Cepheus A HW2. \emph{Astrophys. J.} \textbf{732}, L27 (2011)

\item{} 42. Acknowledgments: The National Radio Astronomy Observatory is a facility of the National Science Foundation, operated under cooperative agreement by Associated Universities, Inc. CC-G acknowledges support by 
UNAM - Dirección General de Asuntos del Personal Académico - Programa de Apoyo a Proyectos de Investigación e Innovación Tecnológica grant number IA101214. GA, JFG, and JMT acknowledge support from Ministerio de Ciencia e Innovaci\'on (Spain) grant AYA2011-30228-C03 (co-funded with Fondo Europeo de Desarrollo Regional funds). JC and SC acknowledge the support of Direcci\'on General de Asuntos del Personal Acad\' emico, UNAM and CONACyT (M\'exico). WV acknowledges support from the European Research Council through consolidator grant 614264.
The ICC (UB) is a CSIC-Associated Unit through the ICE.

\end{itemize}

\noindent \textbf{Supplementary Materials} \\
Materials and Methods \\
A. Observations and data processing \\
B. Spatial alignment of the different continuum bands \\
C. K band continuum in VLA 2 as observed in epochs 1996 and 2014 \\
D. Spectral energy distribution of VLA 2 \\
E. Toroidal environmental density stratification model \\
F. Predictions of radio recombination lines \\
Tables S1 to S4 \\
Figs. S1 to S4\\
References (30-41)\\

\end{document}